\documentclass[conference]{IEEEtran}
\IEEEoverridecommandlockouts
\usepackage{hyperref} 
\usepackage{amsmath,amssymb,amsfonts}
\usepackage{algorithmic}
\usepackage{graphicx}
\usepackage{textcomp}
\usepackage{xcolor}
\usepackage{cite}
\usepackage{booktabs}
 \usepackage{multirow}
\usepackage{pifont}               
\newcommand{\cmark}{\ding{51}}    
\newcommand{\xmark}{\ding{55}}    

\def\BibTeX{{\rm B\kern-.05em{\sc i\kern-.025em b}\kern-.08em
    T\kern-.1667em\lower.7ex\hbox{E}\kern-.125emX}}
\begin{document}

\title{Conan: A Chunkwise Online Network for Zero-Shot Adaptive Voice Conversion
}


\author{
\IEEEauthorblockN{Yu Zhang$^\dagger$}
\IEEEauthorblockA{\textit{Computer Science} \\
\textit{Zhejiang University}\\
Hangzhou, China \\
yuzhang34@zju.edu.cn}
\and
\IEEEauthorblockN{Baotong Tian$^\dagger$}
\IEEEauthorblockA{\textit{Electrical and Computer Engineering} \\
\textit{University of Rochester}\\
Rochester, USA \\
baotong.tian@rochester.edu}
\and
\IEEEauthorblockN{Zhiyao Duan$^\ast$}
\IEEEauthorblockA{\textit{Electrical and Computer Engineering} \\
\textit{University of Rochester}\\
Rochester, USA \\
zhiyao.duan@rochester.edu}
\thanks{$^\dagger$ Equal contribution.}
\thanks{$^\ast$ Corresponding author.}
}

\maketitle

\begin{abstract}
Zero-shot online voice conversion (VC) holds significant promise for real-time communications and entertainment. 
However, current VC models struggle to preserve semantic fidelity under real-time constraints, deliver natural-sounding conversions, and adapt effectively to unseen speaker characteristics.
To address these challenges, we introduce Conan, a chunkwise online zero-shot voice conversion model that preserves the content of the source while matching the speaker representation of reference speech.
Conan comprises three core components: 
1) a Stream Content Extractor that leverages Emformer for low-latency streaming content encoding; 
2) an Adaptive Style Encoder that extracts fine-grained stylistic features from reference speech for enhanced style adaptation; 
3) a Causal Shuffle Vocoder that implements a fully causal HiFiGAN using a pixel-shuffle mechanism. 
Experimental evaluations demonstrate that Conan outperforms baseline models in subjective and objective metrics.
Audio samples can be found at \url{https://aaronz345.github.io/ConanDemo}.
\end{abstract}


\section{Introduction}

Voice Conversion (VC) refers to the task that aims to extract the linguistic content from the source speech and modify its speaker representation to match that of a reference speech~\cite{wang2024streamvoice}. 
The field of VC has seen substantial advances \cite{wang2021vqmivc}. A particularly influential line of work is zero-shot VC, which transfers speaker representations of unseen speakers from reference speech to generated speech, without any speaker-specific fine-tuning~\cite{qian2019autovc}. Most current zero-shot systems typically employ pre-trained feature extraction networks to extract content information from the source speech and combine it with speaker embeddings obtained from the reference speech~\cite{chen2022controlvc}.

The growing need for real-time communication, interactive entertainment, virtual humans, and multimedia production pushes the development of online zero-shot voice conversion systems, and real-time VC systems capable of operating online have been successfully deployed in low-latency scenarios using chunkwise streaming methods \cite{yang2022streamable,yang2024streamvc}. 
However, these methods still fall short of achieving the semantic fidelity, speaker similarity, and naturalness that high-quality, customizable audio experiences require~\cite{zhang2025tcsinger}.
For chunkwise online source speech input, it is essential not only to maintain content extraction and speech synthesis quality under strict online latency constraints, but also to transfer the unseen speaker representation, including global representation (like timbre) and detailed representation (like emotion and prosody) \cite{zhang2025isdrama}.

Currently, zero-shot online VC faces three major challenges:
\begin{itemize}
\item \textbf{Content extraction via chunkwise streaming often compromises quality.} 
Existing VC methods typically employ pretrained feature extraction models to derive linguistic representations, often leveraging self-supervised models such as HuBERT \cite{van2022comparison} or WavLM \cite{chen2022wavlm}. However, the large receptive fields of these pretrained encoders make them unsuitable for online scenarios with minimal or zero lookahead. 
Attempts to train a lightweight causal content encoder via distillation may have fallen short in efficiency and accuracy \cite{yang2024streamvc}. 
Alternatively, some approaches rely on phonetic extraction based on automatic speech recognition \cite{sun2016phonetic}, which incurs additional latency.

\item \textbf{Achieving style transfer in zero-shot scenarios is still challenging.}
Most voice conversion models are not zero-shot and many of them rely on the assumption that the reference voice is accessible for model adaptation~\cite{li21e_interspeech}, which does not always hold in practice. Moreover, many VC approaches focus solely on timbre transfer by applying the source speech’s F0 and energy; This does not capture the speaking style of the reference speaker~\cite{yang2024streamvc}. 
The handful of zero-shot VC methods that do attempt to generalize to unseen speakers still struggle to balance high speaker similarity with naturalness. This shortfall arises from insufficient modeling of speaker representation, preventing effective alignment of speaker information with content \cite{yang2022streamable}.

\item \textbf{Maintaining high quality and naturalness in real-time VC remains difficult.}
In online VC, the system must process source speech chunk by chunk and synthesize generated speech incrementally, which prevents the direct use of conventional VC models and leads to poor naturalness at chunk boundaries \cite{qian2019autovc}. 
Leveraging causality with a constrained look-ahead window is a common approach to address this issue.
Many prior methods employ autoregressive decoders to enforce temporal causality; however, a frame‑by‑frame autoregressive approach often degrades the quality of early synthesized segments and incurs additional computational overhead for the model \cite{wang2024streamvoice}. Additionally, streaming models often convert non-causal vocoders into causal ones by zero-padding, introducing spectral artifacts that degrade quality and efficiency \cite{quamer2024end}.
\end{itemize}

To address these challenges above, we introduce \textbf{Conan}, a \textbf{C}hunkwise \textbf{o}nline \textbf{n}etwork for zero-shot
\textbf{a}daptive voice conversio\textbf{n}.
In zero‐shot online scenarios, Conan preserves the high‐fidelity content of the source speech while matching the unseen speaker representation of reference speech, generating highly natural speech on a chunkwise basis.
To achieve high-quality chunkwise streaming content extraction, we design the \textbf{Stream Content Extractor}, built on an Emformer architecture and trained using content representations extracted offline by HuBERT.
To enable adaptive style transfer, we introduce the \textbf{Adaptive Style Encoder}, which employs clustering-based vector quantization to capture detailed speaker attributes from reference speech and uses an alignment attention mechanism to fuse them with content and timbre information.
For high-quality natural streaming speech synthesis, we adopt a causal-convolutional mel decoder and propose the \textbf{Causal Shuffle Vocoder}, a fully causal HiFiGAN that leverages a pixel-shuffle mechanism to eliminate checkerboard artifacts of transoposed convolution.
Experimental evaluations show that Conan outperforms baseline models in real-time content accuracy, quality, and speaker similarity.
Conan can achieve a latency as low as 37 ms for the conversion on a single A100 GPU without any engineering optimizations.


\section{Related Work}

\subsection{Zero-Shot Voice Conversion}

In practical scenarios, obtaining reference speech during training is often impractical. 
Therefore, zero-shot VC models are essential to capture and control speaker representation by disentangling content from speaker characteristics.
Ebbers et al.~\cite{ebbers2021contrastive} leverage adversarial contrastive predictive coding for fully unsupervised separation of content and speaker.
AutoVC~\cite{qian2019autovc} squeezes out speaker timbre from content embeddings using an information bottleneck, and VQMIVC~\cite{wang2021vqmivc} encodes content via vector quantization, applying mutual information constraints to de-correlate speech components. 
Some approaches obtain speaker representations through a speaker‐verification model and extract content from Automatic Speech Recognition (ASR) posteriorgrams \cite{sun2016phonetic}. 
ControlVC~\cite{chen2022controlvc} uses pretrained encoders to get content and speaker embeddings and applies TD-PSOLA and pitch contour manipulation for time-varying speed and pitch control. 
NANSY~\cite{choi2021neural} trains in a fully self‐supervised manner using wav2vec features alongside Yingram, and LM-VC~\cite{wang2023lm} tokenizes speech into semantic tokens via HuBERT and acoustic tokens via SoundStream. 
Although StreamVoice~\cite{wang2024streamvoice} further advances real‐time separation and recombination of speaker and content using a language model and ASR, current online VC systems still exhibit considerable room for improvement in transferring unseen reference speaker representation and in accurately extracting linguistic content.

\subsection{Online Voice Conversion}

In online VC, the system must process source speech chunk by chunk and synthesize output incrementally, without access to future context. This often degrades naturalness at chunk boundaries. 
For streaming applications, causal processing is a critical design consideration.
Hayashi et al. \cite{hayashi2022investigation} propose a streamable version of the non‐autoregressive sequence‐to‐sequence VC model based on FastSpeech2 \cite{ren2020fastspeech} and NAR‐S2S‐VC \cite{hayashi2021non}, incorporating causal convolutions and self‐attention with causal masks. 
FastS2S‐VC \cite{kameoka2021fasts2s} learns to predict attention distributions from source speech and reference speaker indices alone, guided by a teacher model. 
IBF-VC \cite{chen2023streaming} uses the Intermediate Bottleneck Features (IBFs) to replace Phonetic Posteriorgrams (PPGs) in the ASR encoder to capture more fine-grained prosody information, and applies non-streaming teacher guidance for the timbre leakage problem. 
DualVC \cite{ning2024dualvc} leverages dynamic masked convolution to use the within-chunk future information.
Yang et al. \cite{yang2022streamable} adapt the originally offline VQMIVC model \cite{wang2021vqmivc} for real‐time, chunk‐by‐chunk processing. 
ALO‐VC \cite{wang2023alo} assembles a streaming system composed of a speaker‐verification model, a streamable phonetic‐posteriorgram extractor, and an F0 extractor. 
StreamVC \cite{yang2024streamvc} demonstrates that a lightweight causal convolutional neural network can effectively capture soft speech‐unit representations, while StreamVoice \cite{wang2024streamvoice} introduces autoregressive decoders to enforce temporal causality. 
Quamer et al.~\cite{quamer2024end} convert non-causal vocoders into causal ones by zero-padding.
However, methods relying on frame-by-frame autoregressive structures degrade the quality of early segments and incur extra computational overhead, and zero-padding causal vocoders suffer from artifacts, limiting overall naturalness. 

\subsection{Style Modeling}

Modeling speaking identity remains a central challenge in speech research. Prior approaches have largely relied on pre-trained models to capture only a limited set of styles, like wav2vec 2.0 \cite{baevski2020wav2vec}, HuBERT \cite{hsu2021hubert}, and WavLM \cite{chen2022wavlm}. 
Attentron \cite{choi2020attentron} introduces an attention mechanism to extract speaker representation from reference samples. 
ZSM-SS \cite{kumar2021normalization} proposes a Transformer-based architecture with an external speaker encoder based on wav2vec 2.0. 
Daft-Exprt \cite{zaidi2021daft} employs a gradient reversal layer to improve reference speaker fidelity in style transfer. 
GenerSpeech~\cite{huang2022generspeech} introduces both global and local style adapters to capture diverse speaking identity, while Styler~\cite{lee2021styler} decomposes style into multiple levels of supervision. 
Yang et al.~\cite{yang2022streamable} jointly model speaker identity and global prosody using a GST-based style token network.
Mega-TTS 2 \cite{jiang2024mega} employs vector quantization for prosody encoding combined with a language model for prosody transfer, and NaturalSpeech 3~\cite{ju2024naturalspeech} uses factorized vector quantization to disentangle prosodic features. 
CosyVoice \cite{du2024cosyvoice} integrates x-vectors into a large language model to both disentangle and model prosody. 
Most of these methods focus on offline speech synthesis, while our work addresses the challenge of zero-shot online voice conversion by modeling rich speaking identity and maintaining precise content alignment.

\begin{figure*}[t]
\centering
\includegraphics[width=1.0\textwidth]{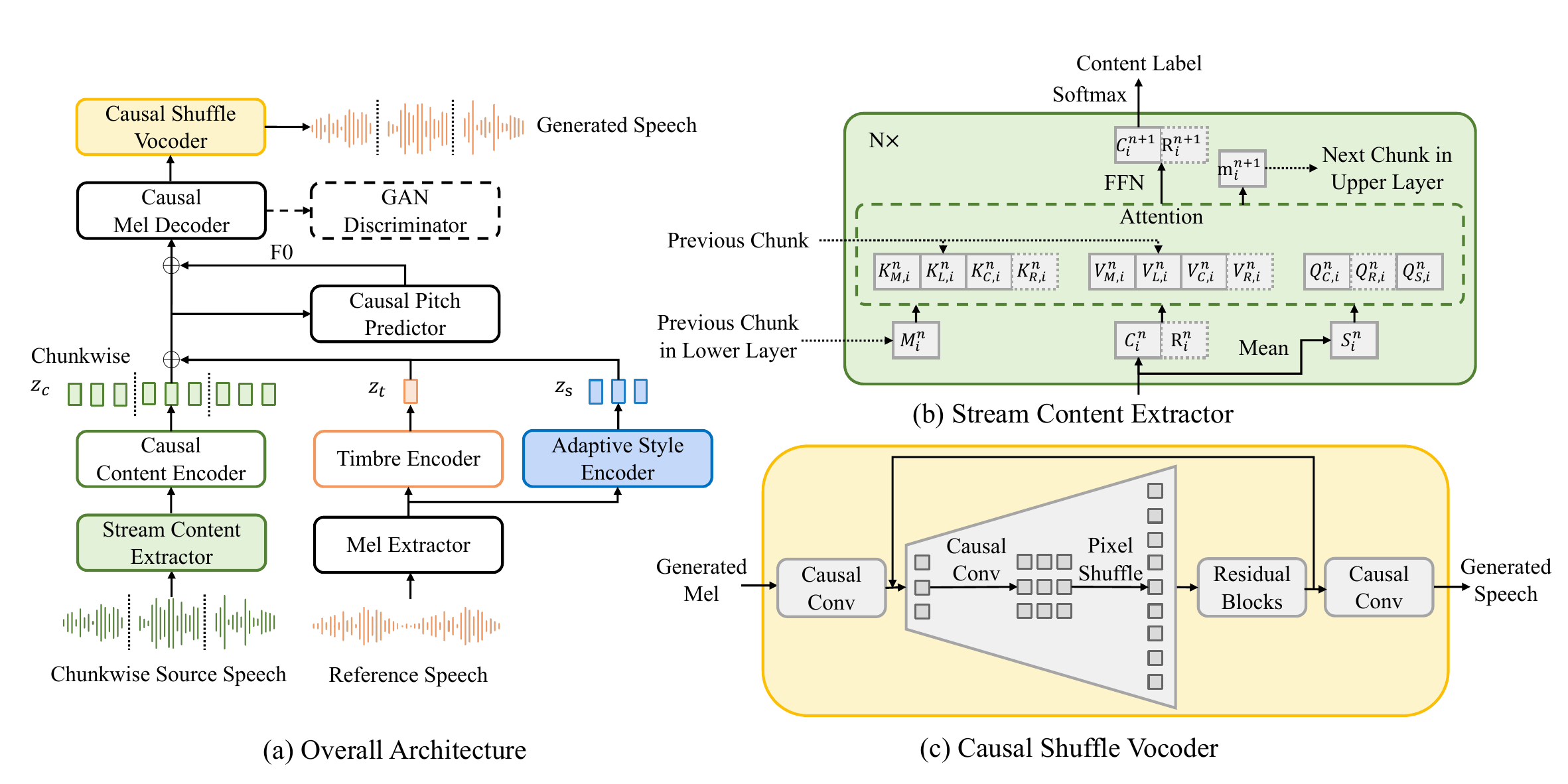}
\caption{The overall architecture of Conan (a). 
Online source speech is fed into the system in chunks, and the synthesized output is likewise produced on a chunkwise basis.
In (b), the size of the right‐context chunk R is configurable; when it is set to zero, the model operates in a chunkwise causal mode.
Here, $C_i^n$ is the content feature of the $i$-th chunk at layer $n$, $R_i^n$ is the right‐context block, $s_i^n$ is the chunk summary, and $M_{i}^n$ is the memory bank from the previous chunks. Attention is computed using $(Q_i^n,K_i^n,V_i^n)$. The output summary $m_{i}^{\,n+1}$ is also obtained and input to the next chunk in the upper layer.
}
 \vspace{-4pt}
\label{fig: arch}
\end{figure*}

\section{Method}


\subsection{Overview}

The architecture of Conan is shown in Figure \ref{fig: arch} (a).
Source speech is input online and fed into the system in chunks. 
In the Stream Content Extractor, it is extracted as content label $z_{c,i}$ for the $i$-th chunk. 
Since the offline HuBERT content representation format is used, each content label corresponds to a 20ms audio segment.
The complete unseen reference speech is passed through a mel extractor to obtain reference mel $m_{rf}$, which is then input to the convolutional timbre encoder to produce a timbre embedding $z_t$.
Additionally, $m_{rf}$ is also fed into the Adaptive Style Encoder to extract chunk-level style representations that may include emotions and prosody. By using attention toward $z_t$ and $z_{c,i}$, we obtain a style representation $z_{s,i}$ that aligns with the $i$-th content chunk.
Next, the causal pitch predictor estimates F0, as explicit pitch modeling greatly aids in conveying speaking styles \cite{lancucki2021fastpitch}.
The causal mel decoder then decodes the content, timbre, style, and pitch embeddings to generate the mel-spectrogram of the $i$-th chunk, which is then converted to the generated speech with high quality and stability via a causal shuffle vocoder. 
This generated speech contains the content of the source speech and the speaker representation of the reference speech.
With sufficient hardware, the processing time for each chunk is shorter than the chunk’s duration, enabling chunkwise, streaming, low-latency, zero-shot online voice conversion.
Note that, except for the Stream Content Extractor’s chunk-input-based design, all other components are frame-level causal models.

\subsection{Stream Content Extractor}

Content accuracy is a critical determinant of VC system quality. However, existing approaches based on pre-trained models or ASR often rely on offline batch processing or incur prohibitive computational costs in chunkwise processing, rendering them unsuitable for streaming VC. Thus, content extractors of online VC must balance efficiency and accuracy. Emformer’s chunk‐based incremental attention mechanism uses cached contextual information from previous chunks to maintain context continuity and capture long‐range dependencies. It preserves high content accuracy while substantially reducing latency, offering an efficient and practical solution for real‐time content extraction from the source utterance.

Our training objective is to distill the content representations extracted by HuBERT \cite{hsu2021hubert} from the source speech at 20 ms intervals—representations that have been widely adopted and validated in prior work \cite{yang2024streamvc}. As shown in Figure \ref{fig: arch} (b), we base our design on Emformer by partitioning the input into chunks for both attention computation and subsequent network processing. During attention computation, we supply context keys, values, and memory‐bank embeddings extracted from the preceding chunk. The memory bank functions provide prior contextual information, with each encapsulating information from one chunk. 
At layer $n$ and chunk $i$, we compute:
\begin{gather}
Q_i^n = \bigl[\,W_q\,C_i^n,W_q\,R_i^n,W_q\,s_i^n\bigr],
\label{eq:Q}\\[0.5em]
K_i^n = \bigl[\,W_k\,M_{i}^n,K_{L,i}^n,W_k\,C_i^n,W_k\,R_i^n\bigr], 
\label{eq:K}\\[0.5em]
V_i^n = \bigl[\,W_v\,M_{i}^n,V_{L,i}^n,W_v\,C_i^n,W_v\,R_i^n\bigr], 
\label{eq:V}
\end{gather}
\begin{gather}
C_i^{\,n+1} =
\mathrm{FFN}\Bigl(\mathrm{Attn}\bigl(W_q\,C_i^n,K_i^n,V_i^n\bigr)+C_i^n\Bigr),
\label{eq:Cupdate}\\[0.5em]
m_{i}^{\,n+1} = \mathrm{Attn}\bigl(W_q\,s_i^n,K_i^n,V_i^n\bigr),
\label{eq:mupdate}\\[0.5em]
z_{c,i} = 
\arg\max_{1 \le j \le J}\Bigl[\,
  \mathrm{Softmax}\bigl(U\,C_i^{(N)}\bigr)
\Bigr]_{(j,t)}
.
\label{eq:zci}
\end{gather}
Here, $Q_i^n$, $K_i^n$, $V_i^n$ refer to the query, key, and value of the $i$-th chunk at layer $n$, $1<n<N$, 
$C_i^n$ is the content feature, $R_i^n$ is the right‐context block, $s_i^n$ is the chunk summary computed by mean pooling, and $M_{i}^n$ is the memory bank from the previous chunk. 
Attention is computed using $(W_q C_i^n,K_i^n, V_i^n)$, the output is added residually to $C_i^n$, and fed into a feedforward network $\mathrm{FFN}$. The summary $m_{i}^{\,n+1}$ is obtained by a similar Attention, forming part of $M_{i}^{n+1}$. Finally, $z_{c,i}$ is produced by projecting $C_i^{(N)}$ with $U$, applying Softmax over $J$ classes per frame $t$, and selecting the highest‐probability label for each frame. All layers remain causal by caching the memory bank $M_{i}^n$, the key $K_{L,i}^n$, and the value $V_{L,i}^n$ from previous segments, enabling low‐latency, zero‐shot online content extraction.

In this attention mechanism, keys, values, and memory‐bank vectors are employed to capture contextual dependencies between data chunks. By flexibly adjusting the size of the right context, one can balance performance against latency; setting it to zero enforces strictly causal behavior. The sizes of the left context and memory buffer determine the overall receptive field, and through our Stream Content Extractor, we achieve efficient, high-quality streaming content extraction.

\subsection{Adaptive Style Encoder}

\begin{figure}[t]
\centering
\includegraphics[width=1.0\linewidth]{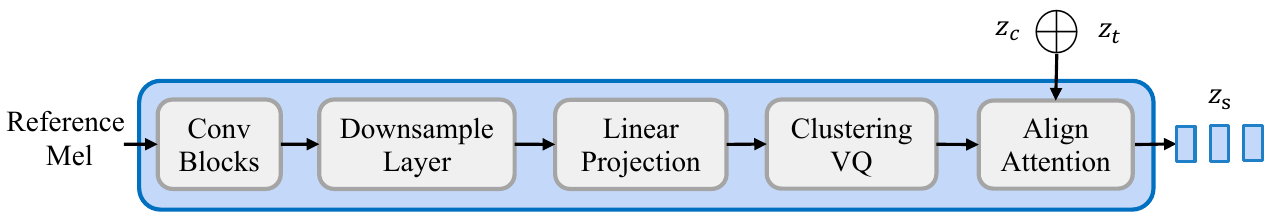}
\caption{The architecture of Adaptive Style Encoder.}
\label{fig: arch2}
 \vspace{-4pt}
\end{figure}

To comprehensively capture and transfer speaker representation from reference speech, we introduce the Adaptive Style Encoder. 
As shown in Figure \ref{fig: arch2}, reference mel-spectrogram $m_{rf}$ is initially refined through convolutional blocks. 
Subsequently, a downsampling layer pools the output into chunk-level representations. In our experiments, we set the chunk size to 80 ms to incorporate styles, which may include emotion and prosody. 

Next, we employ a linear projection to map the outputs $z$ into a low-dimensional latent space for efficient code index lookup \cite{zhang2024tcsinger}. The Clustering Vector Quantization (CVQ) layer \cite{zheng2023online} extracts style representations from these latent inputs, establishing an information bottleneck that effectively filters out non-style information \cite{zhang2025versatile}. 
By applying linear projection and the CVQ mechanism, we capture detailed speaker representation features except global representation and content \cite{huang2022generspeech}.

Compared to traditional VQ \cite{van2017neural}, CVQ employs a dynamic initialization strategy during training. 
This ensures that less frequently or unused code vectors are also updated as frequently used ones do, effectively addressing the codebook collapse issue \cite{zheng2023online}.
We train the CVQ layer using a combination of the VQ loss and contrastive loss, defined as:
\begin{equation}
\begin{aligned}
\mathcal{L}_{CVQ} = &\|sg[z] - e\|^2_2 + \beta \|z - sg[e]\|^2_2+\mathcal{L}_{Contras},
\end{aligned}
\label{eq: vq}
\end{equation}
where $e$ is the selected code from the codebook, $\text{sg}(\cdot)$ is the stop-gradient operator, and $\beta$ is commitment loss hyperparameter. 
For code $e$ in the codebook, we directly select the closest $z^+$ as the positive pair and sample other farther $z^-$ as negative pairs. 
Therefore, the contrastive loss is defined as:
\begin{equation}
\begin{aligned}
&\mathcal{L}_{Contras}=-\log\frac{e^{\text{sim}(e,z^+)}}{e^{\text{sim}(e,z^+)}+\sum_{i=1}^N e^{\text{sim}(e,z_i^-)}}.    
\end{aligned}
\end{equation}

After generating the style embedding, we adaptively align it with the content, taking into account the influence of timbre on style selection, by matching it to $z_{ct,i}$, the concatenation of \(z_{c,i}\) and \(z_t\) \cite{zhang2024stylesinger}.
For this purpose, we introduce the Align Attention module, incorporating the Scaled Dot-Product Attention mechanism \cite{vaswani2017attention}. 
Before inputting the detailed style embedding into the attention module, we also add positional encoding embeddings. 
Within this module, $z_{ct,i}$ serves as the query, while the style embedding functions as both key and value. 
Finally, we get the detailed speaker representation $z_{s,i}$.

\subsection{Causal Shuffle Vocoder}

Enforcing online VC causality through frame-by-frame autoregressive methods degrades the quality of early generated segments and incurs extra computational overhead, while using zero-padding in vocoders leads to spectral artifacts.
To address this, after generating the mel-spectrogram using a causal mel decoder composed of causal convolutions, we build Causal Shuffle Vocoder. 
Based on high-quality HiFi-GAN \cite{kong2020hifi}, we replace standard convolutions with causal convolutions and implement upsampling via pixel shuffle, resulting in a strictly causal vocoder with high audio quality.

As illustrated in Figure \ref{fig: arch} (c), the reference mel-spectrogram first passes through a single causal convolutional layer, producing an initial feature map of shape $(C, T_0)$, where $C$ and $T_0$ represent the channel and time dimensionality, respectively. The network then comprises $N$ upsampling stages. At the $n$-th stage, a causal convolution projects the channel dimension from $C$ up to $r_{n}C$, yielding an intermediate feature map of shape $(r_{n}C, T_{n-1})$. Next, pixel shuffling rearranges these $r_{n}C$ channels into a new tensor of shape $(C, r_{n}T_{n-1})$, where $r_{n}T_{n-1}=T_{n}$, to achieve temporal upsampling. Because pixel shuffling only reorders all channels without introducing additional information from future frames, the network’s process is frame-level causal. By carefully designing the upsampling factors, the final temporal resolution becomes $T_{0}\prod_{n=1}^{N}r_{n}$, thereby recovering the original audio sampling rate.

Immediately following each upsampling stage is a parallel residual block, which refines harmonic and transient details while preserving causality. Within each block, the input and its output are combined via a skip connection, allowing the network to learn only the incremental details rather than the entire spectral envelope. This design ensures fine-grained correction of the spectrum without ever accessing future frames. The outputs of all parallel residual branches at a given scale are averaged before being forwarded to the next upsampling layer.
After the last upsampling layer and its associated residual block, a final causal convolution then maps the channel dimension to 1, producing the generated speech.

By combining causal convolution with pixel shuffling, the waveform produced at each frame depends exclusively on past mel-spectrogram samples.
Causal convolutions pad solely on the left, avoiding spectral distortions from symmetric or right‐side padding.
Pixel shuffling achieves temporal upsampling by simply reorganizing feature‑map channels into a finer time grid, rather than by learning interpolation kernels that span future frames. Because it never blends or spreads information unevenly across time, it inherently avoids the checkerboard artifacts typical of transposed convolutions \cite{odena2016deconvolution}. 
In our strictly causal setup, this means we can match the audio fidelity of HiFi‑GAN while operating online with minimal latency.

\subsection{Training and Inference Procedures}

\begin{figure}[t]
    \centering
    \includegraphics[width=1.0\linewidth]{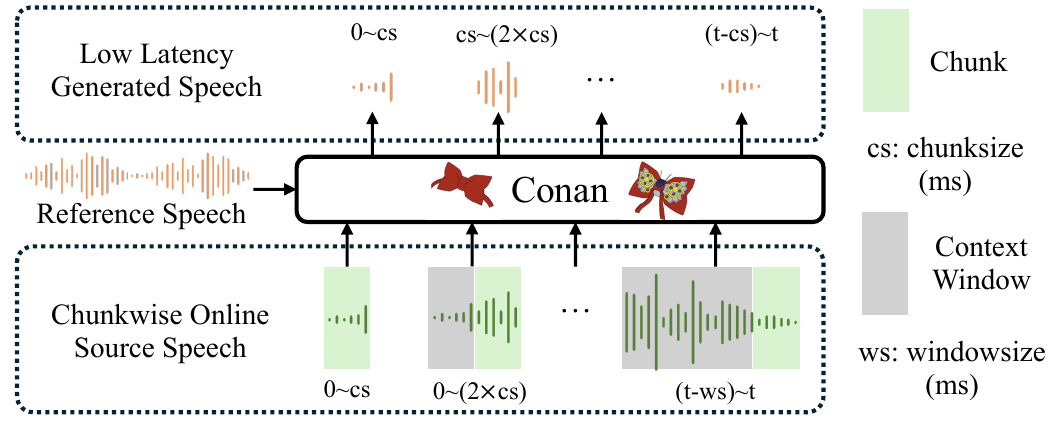}
    \caption{The chunkwise online inference procedure of Conan. 
    Here, we illustrate the causal setting without utilizing any right context.}
    \label{fig: intro}
 \vspace{-4pt}
\end{figure}

\noindent \textbf{Training Procedure.} 
First, for Stream Content Extractor, we use the cross-entropy between the predicted content label and the label extracted by HuBERT. 
Then, the total loss for training the main model comprises $\mathcal{L}_{\mathrm{CVQ}}$ for the adaptive style encoder; $\mathcal{L}_{\mathrm{pitch}}$, the MSE between the ground truth and the predicted $f_{0}$; $\mathcal{L}_{\mathrm{mae}}$ and $\mathcal{L}_{\mathrm{ssim}}$, the MAE and SSIM losses between the predicted and ground truth mel‐spectrograms; and $\mathcal{L}_{\mathrm{GAN}}$, the LSGAN‐style adversarial loss \cite{mao2017least}, whose objective is to minimize the distribution distance between predicted and ground truth mel‐spectrograms. 
Finally, the loss for Causal Shuffle Vocoder follows HiFiGAN \cite{kong2020hifi}, including the GAN loss, mel loss, and feature‐map loss.

\noindent \textbf{Inference Procedure.} 
As shown in Figure~\ref{fig: intro}, we first feed the entire reference speech into the model to provide timbre and stylistic information. During chunkwise online inference, we wait until the input reaches a predefined chunk size before passing it to the model. Because our generation speed for each chunk is faster than the chunk’s duration, online generation becomes possible. To ensure temporal continuity, we employ a sliding context window strategy. At each generation step, we not only input the source speech of the current chunk but also include the preceding context. From the model’s output, we extract only the segment for this chunk. 
As the context covers the receptive field, consistent overlapping segments can be generated, ensuring smooth transitions at chunk boundaries.

\section{Experiments}

\subsection{Experimental Setup}
\noindent \textbf{Dataset.}
We randomly use the source and reference speakers in the train subset of the LibriTTS dataset \cite{zen2019libritts}, which contains 586 hours of English speech from 2,456 speakers. 
All speech is downsampled to 16 kHz. 
To evaluate Conan’s performance in zero‐shot scenarios, we include the VCTK dataset, which comprises 44 hours of clean speech from 110 English speakers. 
For a fair comparison, we reference the StreamVC \cite{yang2024streamvc} data‐partitioning strategy.
We use the test‐clean subset of LibriTTS as the source speech and then randomly select 6 unseen speakers from the VCTK dataset to form the test set.

\noindent \textbf{Implementation Details.}
We set the sample rate to 16 kHz, the window size to 1024, the hop size to 320, and the number of mel bins to 80 to derive mel-spectrograms from raw waveforms.
We set the chunk size to 80 ms and employ a 6-layer Emformer with 2 right context chunks as the \textbf{full setting}.
For the \textbf{fast setting}, the chunk size is 20 ms, and a 3-layer Emformer without right context chunks is employed.
The default size of the codebook for CVQ is 128.

\noindent \textbf{Training Details.}
We train our model using 1 NVIDIA A100 GPU. 
The Adam optimizer is used with $\beta_1 = 0.9$ and $\beta_2 = 0.98$. 
The Stream Content Extractor trains for 80k steps, the main model trains for 160k steps, and the Causal Shuffle Vocoder trains for 600k steps to reach convergence.

\noindent \textbf{Evaluation Details.}
We objectively and subjectively evaluate our model based on three criteria: \textbf{real‐time content accuracy, quality, and speaker similarity}. Content accuracy is measured by Word Error Rate (WER) and Character Error Rate (CER) using a HuBERT‐Large ASR model.
Speaker similarity (SIM) is assessed with Resemblyzer. 
For subjective metrics, we conduct mean opinion score (MOS) and comparative mean opinion score (CMOS) evaluations. Both metrics are rated on a 1–5 scale and reported with 95\% confidence intervals. Specifically, we use MOS‑Q to judge quality (clarity and naturalness) and MOS‑S to assess speaker similarity (timbre and styles) \cite{zhang2024gtsinger}. For CMOS‑Q and CMOS‑S, listeners compare pairs of speech samples produced by different systems. We randomly selected at least 20 pairs with more than 6 unseen speakers for each subjective evaluation, with at least 15 listeners per pair. 
All participants were fairly compensated and provided consent for the use of their responses in scientific research.

\subsection{Main Results}

\begin{table*}[t]
\caption{Evaluation results of Conan and baseline models, including both objective (WER, CER, SIM) and subjective (MOS-S, MOS-Q) evaluations.  
A tick (\cmark) in the \textbf{Online} column indicates streaming/online processing.}
\label{tab: main}
\centering
\small
\scalebox{1}{
\begin{tabular}{llcccccc}
\toprule
\multirow{2}{*}{\bfseries Category} &
\multirow{2}{*}{\bfseries Method} &
\multirow{2}{*}{\bfseries Online} &
\multicolumn{2}{c}{\bfseries Content Accuracy} &
\multicolumn{2}{c}{\bfseries Speaker Similarity} &
\bfseries Quality \\
\cmidrule(lr){4-5} \cmidrule(lr){6-7} \cmidrule(lr){8-8}
 & & & WER $\downarrow$ & CER $\downarrow$ & SIM $\uparrow$ & MOS-S $\uparrow$ & MOS-Q $\uparrow$ \\
\midrule
\multirow{3}{*}{Data} & Source & -- & 5.41\% & 2.43\% & — & — & 3.96 $\pm$ 0.08 \\
 & Source (Vocoder) & -- & 5.56\% & 2.67\% & — & — & 3.94 $\pm$ 0.15 \\
 & Reference (Vocoder) & -- & — & — & 95.81\% & 4.43 $\pm$ 0.14 & — \\
\midrule
\multirow{5}{*}{Baseline} 
 & VQMIVC \cite{wang2021vqmivc}           & \xmark & 56.73\% & 36.18\% & 62.54\% & 3.76 $\pm$ 0.05 & 3.81 $\pm$ 0.07\\
 & BNE-PPG-VC \cite{tang2023any}          & \xmark & 12.29\% & 4.87\%  & 79.06\% & 3.90 $\pm$ 0.13 & 3.86 $\pm$ 0.14\\
 & Diff-VCTK \cite{liu2021any}            & \xmark & 10.58\% & 5.43\%  & 81.47\% & 3.95 $\pm$ 0.07 & 4.05 $\pm$ 0.04\\
 & QuickVC \cite{guo2023quickvc}          & \xmark & 7.14\%  & 3.10\%  & 77.39\% & 3.84 $\pm$ 0.04 & 3.99 $\pm$ 0.09\\
 & StreamVC \cite{yang2024streamvc}       & \cmark & 6.22\% & \textbf{2.17\%} & 77.81\% & 3.86 $\pm$ 0.05 & 4.01 $\pm$ 0.13\\
\midrule
\multirow{2}{*}{Ours} 
 & Conan (Full)                           & \cmark & \textbf{6.02\%} & 2.83\% & \textbf{85.71\%} & \textbf{4.02 $\pm$ 0.06} & \textbf{4.06 $\pm$ 0.04}\\
 & Conan (Fast)                           & \cmark & 10.72\% & 4.74\% & 78.26\% & 3.85 $\pm$ 0.13 & 3.99 $\pm$ 0.11\\
\bottomrule
\end{tabular}}
\end{table*}
We use several popular voice‐conversion models, including VQMIVC \cite{wang2021vqmivc}, BNE‐PPG‐VC \cite{tang2023any}, Diff‐VCTK \cite{liu2021any}, QuickVC \cite{guo2023quickvc}, and StreamVC \cite{yang2024streamvc} as baselines. 
Since StreamVC is not open-sourced, to ensure a fair comparison, we adopt the same dataset split and objective evaluation protocols that they used. 
We rely on the objective metrics reported in their paper and perform subjective assessments based on their demo pages. 

As shown in Table \ref{tab: main}, Conan in the full setting delivers the highest speaker similarity (SIM, MOS-S), indicating that the Adaptive Style Encoder and timbre encoder effectively model the speaker timbre and therefore achieve superior zero-shot similarity. Conan also attains the best quality (MOS-Q), demonstrating that the streaming framework enhances synthesis quality.
For content accuracy, Conan records the lowest WER, while its CER is slightly higher than that of StreamVC. This is mainly because StreamVC reuses the source pitch and energy values, giving it a CER advantage.
Finally, the metrics for Source (Vocoder) and Reference (Vocoder) are both excellent, further highlighting the outstanding online reconstruction capability of the Causal Shuffle Vocoder.

\begin{table}[t]
\caption{
Latency and RTF for Conan’s full and fast settings tested on a single A100 80 GB GPU. 
Latency includes the sum of each module’s delay, plus the chunk size, and the right context.
}
\label{tab: rtf}
\centering
\small
\begin{tabular}{lcccc}
\toprule
\multirow{2}{*}{\bfseries Method} & \multicolumn{2}{c}{\bfseries Full Setting} & \multicolumn{2}{c}{\bfseries Fast Setting}\\ \cmidrule(lr){2-3} \cmidrule(lr){4-5}
& RTF & Latency (ms) & RTF & Latency (ms) \\
\midrule
Content & 0.07 & 5.60 & 0.14 & 2.76 \\
Main & 0.10 & 7.88 & 0.39 & 7.82 \\
Vocoder & 0.08 & 6.21 & 0.31 & 6.29 \\
\midrule
Overall & 0.25 & 139.71 & 0.74 & 36.87 \\
\bottomrule   
\end{tabular}
\end{table}

As shown in Table \ref{tab: rtf}, in the fast setting, Conan uses a 20 ms chunk size to perform strictly causal, chunkwise online voice conversion with only 37 ms of latency, yet still achieves high performance metrics. 
Under Conan’s full setting with 80 ms chunk size and 2 right context chunks, we achieve high quality while still meeting real‐time latency constraints.

\subsection{Ablation Study}

\begin{table}[t]
\caption{
Evaluation results for the ablation study.
SCE denotes Stream Content Encoder, ASE is Adaptive Style Encoder, and CSV represents Causal Shuffle Vocoder.
}
\label{tab: abl}
\centering
\small
\begin{tabular}{lccc}
\toprule
\bfseries{Setting} & WER$\downarrow$ & CMOS-Q $\uparrow$ & CMOS-S $\uparrow$ \\
\midrule
Conan & 6.02\% & 0.00  & 0.00 \\
\midrule
w/o SCE & 26.92\% & -0.33 & -0.18\\
w/o Right & 8.57\% & -0.19 & -0.07\\
w/o ASE  & 6.53\%  & -0.17 & -0.19 \\
w/o CSV  & 7.06\%  & -0.21 & -0.05\\
\bottomrule   
\end{tabular}
\end{table}

As shown in Table \ref{tab: abl}, we conduct ablation studies to demonstrate the effectiveness of various design choices within the full setting. When the original HuBERT is used instead of the Stream Content Encoder, we observe a substantial deterioration in both synthesis quality and WER, underscoring the incompatibility of HuBERT with streaming scenarios. Furthermore, when right‐context chunks are omitted, all metrics worsen, indicating that the size of the lookahead can be adjusted as a trade‐off between efficiency and quality.
Furthermore, removing the Adaptive Style Encoder results in a marked decrease in MOS-S, confirming that our approach effectively captures speaker representation. 
Finally, substituting the Causal Shuffle Vocoder with a zero-padding causal HiFi-GAN leads to a decline in MOS-Q, as the zero padding causal HiFi-GAN causes a decline in synthesis quality.

\section{Conclusion}

In this paper, we introduce Conan, a streaming voice conversion system that preserves the linguistic content of a source utterance while adopting the speaker information of a reference utterance. 
We design Streaming Content Extractor, Adaptive Style Encoder, and Causal Shuffle Vocoder for better zero-shot online VC performance.
Experimental results show that Conan surpasses existing baseline approaches, achieving better performance according to subjective and objective metrics.

\section*{Acknowledgment and Disclaimer}
Supported by the Intelligence Advanced Research Projects Activity (IARPA) via Department of Interior/Interior Business Center (DOI/IBC) contract number 140D0424C0066. The U.S. Government is authorized to reproduce and distribute reprints for Governmental purposes notwithstanding any copyright annotation thereon.

The views and conclusions contained herein are those of the authors and should not be interpreted as necessarily representing the official policies or endorsements, either expressed or implied, of IARPA, DOI/IBC, or the U.S. Government.

\bibliographystyle{IEEEtran}  
\bibliography{custom}   

\end{document}